\documentclass[twocolumn,preprintnumbers,amsmath,amssymb]{revtex4} 

\usepackage{amssymb}
\usepackage{mathrsfs}

\usepackage{hyperref}

\usepackage{graphicx}
\usepackage{dcolumn}
\usepackage{bm}
\usepackage{mathrsfs}
\usepackage{multirow}   
 
\usepackage{atlasphysics} 

\usepackage{lldefs} 

\begin{document}

\begin{flushright}
CERN-PH-EP-2011-037; submitted to Physics Letters B
\end{flushright}

\title{Search for high mass dilepton resonances in $pp$ collisions at $\sqrt{s}=7$~TeV with the ATLAS experiment}
       
\author{The ATLAS Collaboration}

\begin{abstract}
  This article presents a search for high mass  \ee\ or \mumu\ resonances in $pp$ collisions  at $\rts = 7$~\TeV\ at the LHC. 
The data were recorded by the ATLAS experiment during 2010 and correspond to a total integrated luminosity of $\sim 40$~\ipb. 
No statistically significant excess above the Standard Model expectation is observed in the search region of dilepton invariant mass above 110 GeV.
 Upper limits at the 95\% confidence level are set on the cross section times branching ratio of \zp\ resonances decaying to 
 dielectrons and dimuons as a function of the resonance mass. 
A lower mass limit of 1.048~TeV on the Sequential Standard Model \zp\ boson is derived, as well as mass limits on \zstar\ and \esix-motivated \zp\ models.

\end{abstract}


\maketitle


A search for high mass resonances decaying into \ee\ or \mumu\ pairs 
is presented based on an analysis of 7~\TeV\ $pp$ collision data recorded with the ATLAS detector~\cite{atlas:detector}.
Among the possibilities for such resonances, this article focuses on new heavy neutral gauge bosons (\zp, \zstar)~\cite{Langacker:2008yv,Erler:2009jh,wzstar};
other hypothetical states like a Randall-Sundrum spin-2 graviton~\cite{RS} or a spin-1 techni-meson~\cite{Lane:1989ej} are not discussed here, 
though this analysis is also sensitive to them.

The benchmark model for \zp\ bosons is the Sequential Standard Model (SSM)~\cite{Langacker:2008yv}, 
in which the \zp\ (\zpssm) has the same couplings to fermions as the \z\ boson.
A more theoretically motivated model is the Grand Unification
  model in which the \esix\ gauge group is broken into \su5\ and two additional \u1\ groups~\cite{London:1986dk}. 
The lightest linear combination of the corresponding two new neutral gauge bosons, $\zp _\psi$ and $\zp _\chi$, is considered the \zp\ candidate:
$\zp (\te6 ) = \zp _\psi \cos{\te6 } + \zp _\chi \sin{\te6 }$, 
where $0\leq \te6 < \pi$ is the mixing angle between the two gauge bosons. 
The pattern of spontaneous symmetry breaking and the value of \te6\ determines the \zp\ couplings to fermions;
six different models~\cite{Langacker:2008yv,London:1986dk} lead to the specific \zp\ states named \zppsi, \zpN, \zpeta, \zpI, \zpsq\ and \zpchi\ respectively.
Because of different couplings to $u$ and $d$ quarks, the ranking of the production cross sections of these six states 
is different in $p\bar p$ and $pp$ collisions.  
In this search, the resonances are assumed to have a narrow intrinsic width, comparable to the contribution from the detector mass resolution.
The expected intrinsic width of the \zpssm\  as a fraction of the mass is 3.1\%, while
for any \esix\ model the intrinsic width is predicted to be between 0.5\% and 1.3\%~\cite{Dittmar:2003ir}.

Production of a \zstar\ boson~\cite{wzstar,wzstar_motivate} could also be detected in a dilepton resonance search. 
The anomalous (magnetic moment type) coupling of the \zstar\ boson leads to kinematic distributions different from those of the \zp\ boson.
To fix the coupling strength, a model with quark-lepton
universality, and with the total \zstar\ decay width equal to that of
the \zpssm\ with the same mass, is adopted~\cite{wzstar_refmod, wzstar_refmod2}.

Previous indirect and direct searches have set constraints 
on the mass of \zp\ resonances~\cite{resonanceReview,Abbiendi:2003dh,Abdallah:2005ph,Achard:2005nb,Schael:2006wu}. 
The \zpssm\ is excluded by direct searches at the Tevatron with a mass lower than  
1.071~TeV~\cite{Abazov:2010ti,CDF:Zpmumu}.
The large center of mass energy of the LHC provides an opportunity to search for \zp\ resonances with comparable sensitivity 
using the 2010 $pp$ collision data.  
CMS has very recently excluded a \zpssm\ with a mass lower than 1.140~TeV~\cite{CMS_Zp2010}.

The three main detector systems of ATLAS~\cite{atlas:detector} used in this analysis are the inner tracking detector, the calorimeter, 
and the muon spectrometer. 
Charged particle tracks and vertices are reconstructed with the inner detector (ID) which consists of silicon pixel, silicon strip, 
and transition radiation detectors
covering the pseudorapidity range $|\eta| < 2.5$~\footnote{ATLAS uses a right-handed coordinate system with its origin at the nominal interaction point (IP) in the centre of the detector and the $z$-axis along the beam pipe. The $x$-axis points from the IP to the centre of the LHC ring, and the $y$ axis points upward. Cylindrical coordinates $(r,\phi)$ are used in the transverse plane, $\phi$ being the azimuthal angle around the beam pipe. The pseudorapidity is defined in terms of the polar angle $\theta$ as $\eta=-\ln\tan(\theta/2)$.}.
It is immersed in a homogeneous 2~T magnetic field provided by a superconducting solenoid.
The latter is surrounded by a finely-segmented, hermetic calorimeter that covers $|\eta| < 4.9$
and provides three-dimensional reconstruction of particle showers using lead-liquid argon sampling for the electromagnetic
compartment followed by a hadronic compartment which is based on iron-scintillating tiles sampling in the central region
and on liquid argon sampling with copper or tungsten absorbers for $|\eta| > 1.7$.
Outside the calorimeter, there is a muon spectrometer with air-core toroids providing a magnetic field. 
Three sets of drift tubes or cathode strip chambers provide precision ($\eta$) coordinates for momentum measurement in the region $|\eta |< 2.5$.
Finally, resistive-plate and thin-gap chambers provide muon triggering capability.


The data sample used in this analysis was collected during 2010. 
Application of detector and data quality requirements leads to an available integrated
luminosity of 39~\ipb\ and 42~\ipb\  for the electron and muon analyses respectively.

Triggers requiring the presence of at least one electron or muon above a transverse momentum (\pt) threshold were used to identify
the events recorded for full reconstruction.  The thresholds varied from 14 to 20~\gev\ for electrons
and 10 to 13~\gev\ for muons depending on the luminosity.
The overall trigger efficiency at the \z\ peak is 100\% with negligible uncertainty for dielectron events
and $(98.2\pm 0.3)$\% for dimuon events, for the selection criteria presented below.
The trigger-level bunch-crossing identification of very high transverse energy electron triggers 
relies on a special algorithm implemented in the first-level calorimeter trigger hardware;
its performance was checked with calibration data and the resulting systematic uncertainty on the trigger efficiency is $^{+0}_{-2}$\%.
Collision candidates are selected by requiring a primary vertex  with at least three associated charged particle tracks, consistent
with the beam interaction region.



In the \ee\ channel,
two electron candidates are required with transverse energy $E_{T} > 25$~\gev,
$|\eta| < 2.47$; the region $1.37\leq |\eta|\leq 1.52$ is excluded because it corresponds
to a transition region between the barrel and endcap calorimeters which has degraded energy resolution.
Electron candidates are 
formed from clusters of cells reconstructed in the electromagnetic calorimeter. 
Criteria on the transverse shower shape, the longitudinal leakage into the hadronic calorimeter, and the association to an inner detector
 track are applied to the cluster to satisfy the  \Medium\ electron definition~\cite{atlas:egamma_id_900GeV,Aad:2010yt}.
The electron energy is obtained from the calorimeter measurements and its direction from the associated track.
A hit in the first layer of the pixel detector is required (if an active pixel layer is traversed) to suppress background from photon conversions.
In addition, a fiducial cut removes events with electrons near problematic regions of the electromagnetic calorimeter during the 2010 run,
reducing the acceptance by 6\%.
The two electron candidates are not required to have opposite charge because of possible charge mis-identification either due to bremsstrahlung
or to the limited momentum resolution of the inner detector at very high \pt. 
For these selection criteria, the overall event acceptance  for a $\zpee$ of mass 1~\tev\  is 60\%.

In the \mumu\ channel,
two muon candidates of opposite charge are required, each satisfying $\pt > 25$~\gev.
These muons are required to be within the trigger acceptance of $|\eta| < 2.4$.
Muon tracks are reconstructed independently in both the inner detector and muon spectrometer. 
The momentum is taken from a combined fit to the measurements from both subsystems. To obtain optimal momentum resolution, 
the muons used in this analysis are required to have at least three hits in each of the inner, middle, and outer detectors of the muon system,
and at least one hit in the non-bend plane. 
Residual misalignments of the muon detectors, which could cause a degradation of the momentum resolution, were studied with cosmic rays
and with collision data in which the muons traversed overlapping sets of muon chambers.
The effect of the misalignments, and the intrinsic position resolution, are included in the simulation.
Studies of  muons from \w\ decays verified that the observed momentum spectrum agrees with the simulation up to $\pt=300$~\gev\
above which the event numbers are sparse. 
To suppress background from cosmic rays, the muons are also required to satisfy selections on the impact parameter, $|d_0|<0.2$~mm;
$z$ coordinate with respect to the primary vertex (PV), $|z_0-z({\rm PV})|<1$~mm;
and on the $z$ position of the primary vertex, $|z({\rm PV})|<200$~mm.
To reduce the background from jets, each muon is required to be isolated such that 
 $\Sigma\pt(\Delta R<0.3)/\pt(\mu)<0.05$, where $\Sigma \pt (\Delta R < 0.3)$ is the sum of the \pt\ of the other tracks in a cone
  $\Delta R<0.3$ around the direction of the muon ($\Delta R=\sqrt{(\Delta \eta)^2 + (\Delta \varphi)^2}$).
  The overall event acceptance 
  is 40\% for a \zpmumu\ of mass 1~\tev. The primary
  reason for the lower acceptance compared to the electron channel is the requirement that hits are observed
  in all three layers of muon chambers, which reduces coverage in some regions of \eta. 
  It is expected that this acceptance difference will be recovered in the future.


For both channels, the dominant background originates with 
the $Z/\gamma^*$ (Drell-Yan) process,
which has the same final state as \zp\ or \zstar\ production.
In the \ee\ channel, the second largest background arises from QCD jet production including $b$ quarks 
(referred to below as QCD background); above $\mee=110$~GeV, the next largest backgrounds are \ttbar\ and \wpjet\ events.
In the \mumu\ channel, 
in order of dominance the backgrounds are  Drell-Yan production, followed by \ttbar\
and diboson ($WW$, $WZ$ and $ZZ$) production; the QCD and \wpjet\ backgrounds are negligible.

Expected signal and backgrounds, with the exception of the QCD component,
are evaluated with simulated samples and normalized with respect to one another 
using the highest-order available cross section predictions.
The \zp\ signal and $Z/\gamma^*$ processes are generated with
\pythia\ 6.421~\cite{Sjostrand:2006za} using MRST2007 LO*~\cite{mrst} parton distribution functions (PDF's).
The \zpssm\ was used as the benchmark signal model and this signal sample was generated with \pythia\ using 
Standard Model couplings. 
\zstar\  events are generated with \comphep\ \cite{comphep} 
 using CTEQ6L1~\cite{Pumplin:2002vw} PDF's followed by \pythia\ for parton showering and underlying event generation.
The diboson processes are generated with \herwig~6.510~\cite{herwig,Corcella:2002jc} using MRST2007 LO* PDF's.
The \wpjet\ background is generated with \alpgen~\cite{Mangano:2002ea} and
the \ttbar\ background with \mcatnlo~3.41~\cite{mcatnlo}.
For both, \jimmy~4.31~\cite{jimmy} is used to describe multiple parton interactions and
\herwig\ to describe the remaining underlying event and parton showers.
CTEQ~\cite{Pumplin:2002vw} parton distribution functions are used.
For all samples, final state photon radiation is handled by \photos~\cite{fsr_ref} and
the interaction of particles and the response of the detector are carried out using full detector
simulation~\cite{atlas:sim} based on \geant4~\cite{geant}.

The $Z/\gamma^*$ cross section is calculated at next-to-next-to-leading order (NNLO)
using PHOZPR~\cite{Hamberg:1990np} with MSTW2008 parton distribution functions~\cite{mstw}. 
The ratio of this cross section to the leading order cross section can be used to determine a mass dependent QCD K-factor which is 
applied to the results of the leading order simulations. 
The same QCD K-factor is applied to the \zp\ signal. However, the QCD K-factor is not applied to the leading-order \zstar\ cross section
 since the \zstar\ model uses an effective Lagrangian with a different Lorentz structure. 
 Higher order weak corrections (beyond the photon radiation included in the simulation) 
are calculated using \horace~\cite{horace,CarloniCalame:2007cd}, yielding a weak K-factor due to virtual heavy gauge boson loops.
The weak K-factor is not applied to the \zp\ or \zstar\ signal since it is not universal,
but depends on the coupling of the \w\ and \z\ bosons to the \zp\ or \zstar.
The diboson cross section is known to next-to-leading order (NLO) with an uncertainty of 5\%.
The \wpjet\ cross section is calculated at NLO, and rescaled to the inclusive NNLO calculation,
resulting in 30\% uncertainty when at least one parton with $E_{T} > 20$~\gev\ accompanies the \w\ boson.
The \ttbar\ cross section is  predicted at approximate-NNLO, with 10\% uncertainty~\cite{Moch:2008qy,Langenfeld:2009tc,Aliev:2010zk}.
Cross section uncertainties are estimated from PDF error sets
and from variation of renormalization and factorization scales in the cross section calculation.


To estimate the QCD background in the \ee\ sample, a combination of three different techniques is used. 
In the ``reversed electron identification'' technique, a sample of events where both electrons pass the \Loose\ 
electron identification selections~\cite{atlas:egamma_id_900GeV,Aad:2010yt} but fail the \Medium\ selections is used to determine the shape of the 
QCD background as a function of 
invariant mass \mee.
This template shape, and the sum of the Drell-Yan, diboson, \ttbar\ and \wpjet\ backgrounds, 
are fitted to the observed \mee\ distribution to determine the normalization of the QCD contribution.  
In the second technique~\cite{Aad:2010yt}, the isolation distribution for the electrons 
(energy in the calorimeter in a cone of $\Delta R < 0.4$ around the electron track after subtracting the electron cluster energy) 
is fitted to a signal template, corresponding to electrons from either \z\ or \zp/\zstar\ production,
plus a background template; the latter is determined from the data by 
reversing electron identification selections.
The third technique relates, via a matrix inversion, the measured number of events passing \Loose\
or \Medium, plus first-pixel-layer hit, identification requirements for each of the two electrons
 (i.e. four different categories of events) to the true number of real and fake electron combinations 
 in the sample~\cite{atlas:electron_performance,Aad:2010ey}.
To combine the measurements from each of these estimates and obtain the QCD background in the high-\mee\ region, 
a fit  in several bins of \mee\ above 110~\gev\ is performed using 
a power-law function of \mee\ with the parameters being the exponent
and the integral number of events with $\mee>110$~\gev.
The  background in any given region of \mee\ is then obtained from an integral of this function; the corresponding uncertainty is obtained by propagating 
the statistical and systematic uncertainties for each of the background estimation methods.
A small additional systematic uncertainty related to a small bias in the fit for low statistics and  variations when different functions were used
is also taken into account. 
The power law function gives a conservative estimate of the QCD background at very large \mee, as it falls less rapidly than 
other functional forms used to fit dijet invariant mass distributions~\cite{Atlas_First_Dijet}.

  QCD backgrounds in the \mumu\ sample can be produced by pion and kaon decay in flight or from semi-leptonic decays of $b$ and $c$ quarks.
  The former is suppressed by the small decay probability of a high-\pt\ pion or kaon.
 The background from  semi-leptonic decays of $b$ and $c$ quarks is evaluated using the $\Sigma \pt (\Delta R < 0.3) / \pt (\mu)$ isolation variable. 
 A simulated sample of $b \bar{b}$ and
 $c \bar{c}$ events is shown to reproduce the isolation distribution of the muon candidates, after all selection cuts except isolation are applied. This 
 simulated QCD sample is normalized to the data in the region $\Sigma \pt (\Delta R < 0.3) / \pt (\mu) >0.1$, and then used to predict the background passing
 the final selection criterion of $\Sigma \pt (\Delta R < 0.3) / \pt (\mu)  < 0.05$.  A systematic uncertainty of 50\% is assigned to the QCD background 
 to cover the difference between the number of non-isolated muons predicted by the simulation and the number observed in the data.
 
 A direct estimate of background from cosmic rays in the muon channel is obtained 
by observing the rate, and mass distribution, of events satisfying $3<|z_0-z({\rm PV})|<200$~mm or $|d_0|>0.3$~mm. The number 
of events in the final sample is obtained by scaling to the number expected to pass the $|d_0|<0.2$~mm, and
 $|z_0-z({\rm PV})|<1$~mm selection criteria. The total cosmic ray background above $\mmumu=70$~GeV is thus estimated to be
 $0.004 \pm 0.002$ events.

 Finally, while the primary estimate of the $t \bar{t}$ background is taken from  Monte Carlo for both channels as discussed above, 
 a data-driven cross-check of the $t \bar{t}$ background was also performed. 
 The $e \mu$ final state with dilepton invariant mass $>100$~GeV provides an enriched sample of \ttbar\ fully leptonic events.
  After correcting for relative efficiencies,
 it provides a direct estimate from data of the $\ttbar \to \ee$, \mumu\ backgrounds. The results, which have
 relatively large statistical uncertainties due to the limited number of events, are in good agreement with the Monte Carlo prediction.


The observed invariant  mass distributions, \mee\ and \mmumu, are compared to the expectation of the SM backgrounds. To make this 
comparison, the sum of the Drell-Yan, $t \bar{t}$, diboson and \wpjet\ backgrounds (with the relative contributions fixed according to
the respective cross sections) is scaled such that when added to the data-driven QCD background, 
the result agrees with the observed number of data events in the 70 - 110~\gev\ mass interval. The advantage of this approach is
that the uncertainty on the luminosity, and any mass independent uncertainties in efficiencies, cancel between the \zp/\zstar\ and the \z\
in the limit computation presented below. 
The integrated Drell-Yan cross section at NNLO above a generator-level dilepton invariant mass of 60~GeV is $(0.989 \pm 0.049)$~nb.

\begin{figure}[!htb]
  \hspace*{-0.6cm}
  \includegraphics[width=\columnwidth]{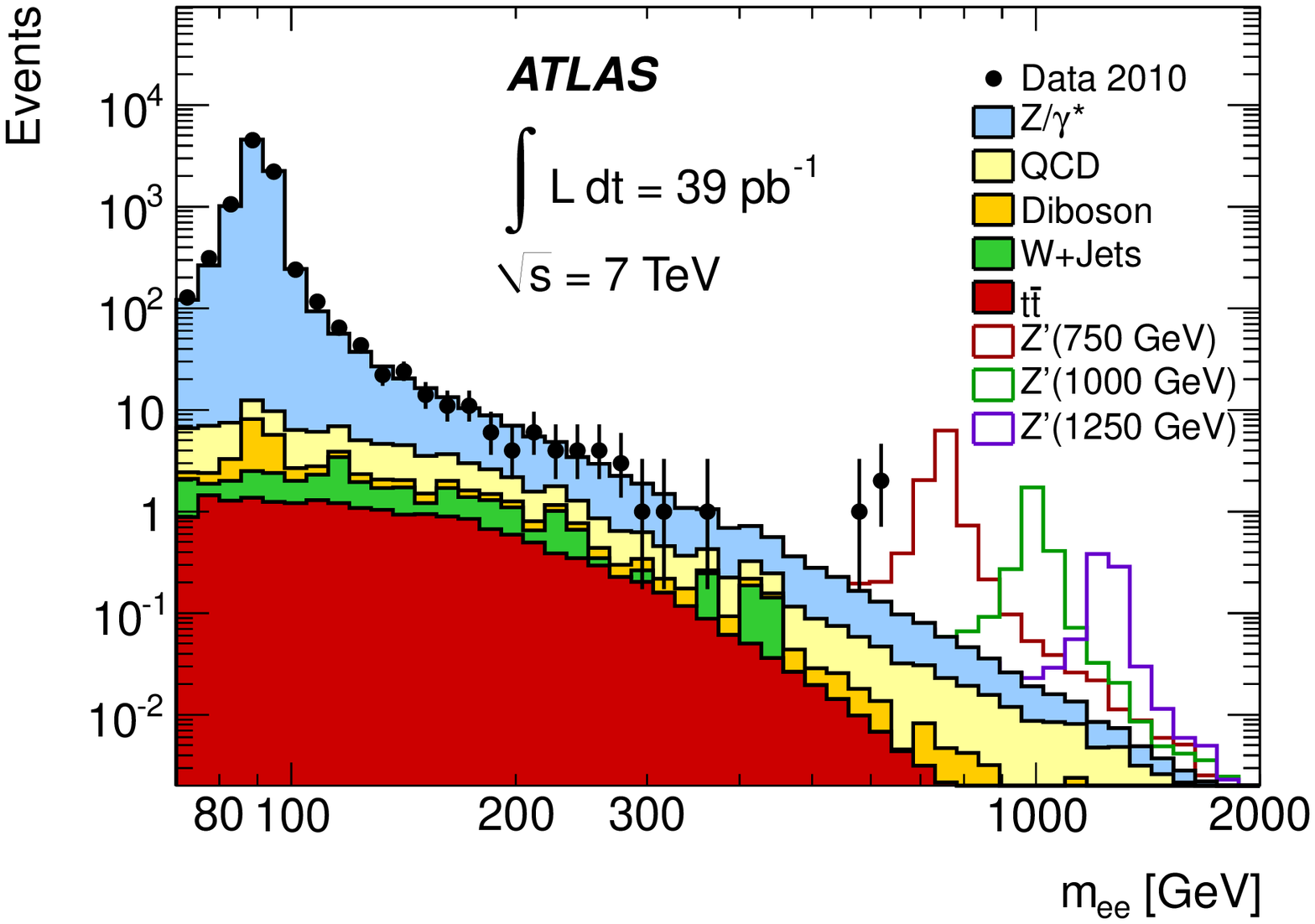}
  \caption{Dielectron invariant mass (\mee) distribution after final selection, compared to the stacked sum of all expected backgrounds,
  with three example \zpssm\ signals overlaid. The bin width is constant in $\log \mee$.}
  \label{fig:init_res3}
\end{figure}

Figure~\ref{fig:init_res3} presents the invariant mass (\mee) distribution after final selection
while Table~\ref{tab:dielectronTable_syst} shows the number of data events and estimated backgrounds in bins of reconstructed \ee\ invariant mass.
The dielectron invariant mass distribution is well described by the prediction from SM processes.

\begin{figure}[!htb]
\hspace*{-0.6cm}
\includegraphics[width=\columnwidth]{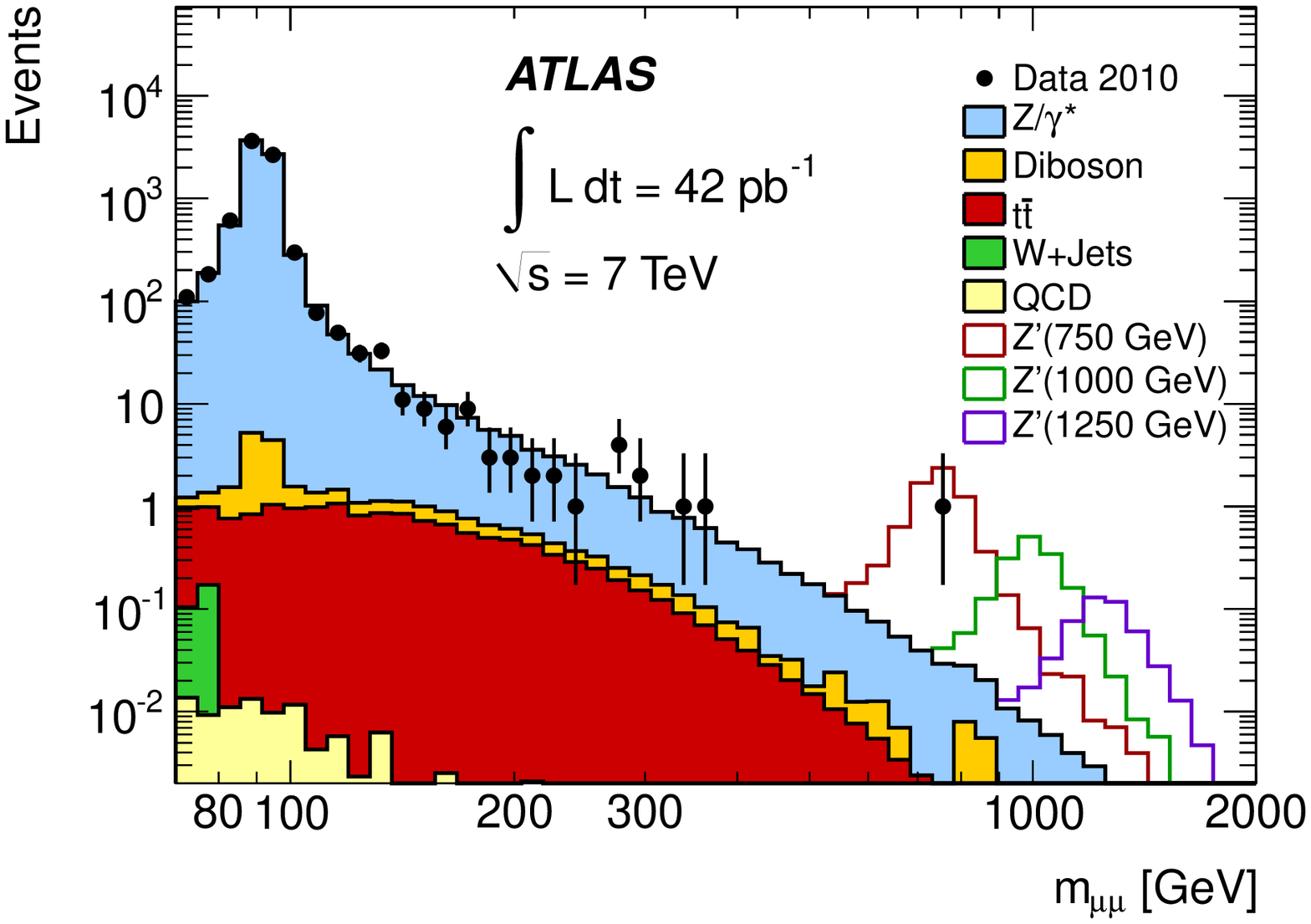}
  \caption{Dimuon invariant mass (\mmumu) distribution after final selection, compared to the stacked sum of all expected backgrounds,
  with three example \zpssm\ signals overlaid. The bin width is constant in $\log \mmumu$. }
  \label{fig:moneyPlot}
\end{figure}

\begin{table}[!htb]
\caption{Expected and observed number of events in the dielectron channel.
  The uncertainties quoted include both statistical and systematic uncertainties.
The systematic uncertainties are correlated across bins and are discussed in the text. 
Entries of 0.0 indicate a value $ < 0.05$.}
\label{tab:dielectronTable_syst}
\centering
\addtolength{\tabcolsep}{+2pt}
\resizebox{\columnwidth}{!}{
\begin{tabular}{c r@{ $\pm$ }r r@{ $\pm$ }r r@{ $\pm$ }r r@{ $\pm$ }r r@{ $\pm$ }r r@{ $\pm$ }r}
\hline
\hline
\mee [GeV]& \multicolumn{2}{c}{\phantom{00}70-110}& \multicolumn{2}{c}{\phantom{00}110-130}& \multicolumn{2}{c}{130-150}& \multicolumn{2}{c}{150-170}& \multicolumn{2}{c}{170-200}\\
\hline
\zgstar & 8498.5& 7.9 & 104.9&3.3 & 36.8&1.3 & 19.4&0.7 & 14.7&0.6 \\
\ttbar  &    8.2& 0.8 &   2.8&0.3 &  2.1&0.2 &  1.7&0.2 &  1.7&0.2 \\
Diboson &   12.1& 0.9 &   1.0&0.2 &  0.7&0.2 &  0.5&0.2 &  0.5&0.1 \\
\wpjet  &    6.0& 1.8 &   3.7&1.2 &  1.2&0.5 &  1.3&0.5 &  1.2&0.4 \\
QCD     &   32.1& 7.1 &   8.4&1.8 &  5.5&0.8 &  3.2&0.6 &  2.8&0.8 \\
\hline
 Total  & 8557.0&10.8 & 120.9&4.0 & 46.4&1.6 & 26.2&1.1 & 20.8&1.1 \\
\hline
Data & \multicolumn{2}{c}{8557} & \multicolumn{2}{c}{131} & \multicolumn{2}{c}{49} & \multicolumn{2}{c}{20} & \multicolumn{2}{c}{18} \\
\hline
\hline
\mee [GeV]& \multicolumn{2}{c}{\phantom{000}200-240}& \multicolumn{2}{c}{\phantom{00}240-300}& \multicolumn{2}{c}{300-400}& \multicolumn{2}{c}{400-800}& \multicolumn{2}{c}{\phantom{00}800-2000 } \\
\hline
\zgstar & 9.5&0.4 & 6.0&0.3 & 3.2&0.1 & 1.6&0.1 & 0.1&0.0 \\
\ttbar  & 1.2&0.1 & 0.9&0.1 & 0.5&0.0 & 0.2&0.0 & 0.0&0.0 \\
Diboson & 0.4&0.1 & 0.3&0.1 & 0.2&0.1 & 0.1&0.1 & 0.0&0.0 \\
\wpjet  & 1.1&0.4 & 0.3&0.1 & 0.2&0.1 & 0.2&0.1 & 0.0&0.0 \\
QCD     & 1.9&0.8 & 1.3&0.7 & 0.8&0.4 & 0.5&0.2 & 0.1&0.1 \\
\hline
 Total  &14.1&1.0 & 8.8&0.7 & 4.8&0.5 & 2.7&0.3 & 0.2&0.1 \\
\hline
Data & \multicolumn{2}{c}{13} & \multicolumn{2}{c}{9} & \multicolumn{2}{c}{3} & \multicolumn{2}{c}{3} & \multicolumn{2}{c}{0} \\
\hline
\hline
\end{tabular}
}
\end{table}
%
%
%
Similarly, Figure~\ref{fig:moneyPlot}, and Table~\ref{tab:backgroundTableMuon_syst} show the results for the \mumu\ sample.
Again, there is good agreement with the prediction from SM processes.
Figures~\ref{fig:init_res3} and \ref{fig:moneyPlot} also display expected \zpssm\ signals for three masses around 1~TeV.
Expected \zstar\ signals (not shown) have a similar shape and approximately 40\% higher cross section.
 Three events in the vicinity of $\mee = 600$~GeV
  and  a single event at $\mmumu = 768$~GeV are observed in the data. 
 The $p$-value which quantifies, in the absence of signal, the probability of observing an excess anywhere 
 in the search region $\mll > 110$~GeV ($\ell=e$ or $\mu$), 
 with a significance at least as great as that observed in the data is evaluated~\cite{bayesianMethod}.
 Since the resulting $p$-values are 5\% and 22\% for the electron and muon channels, respectively, no statistically significant
 excess above the predictions of the SM has been observed.
\begin{table}[!htb]
\caption{Expected and observed number of events in the dimuon channel.
The uncertainties quoted include both statistical and systematic uncertainties. 
The systematic uncertainties are correlated across bins and are discussed in the text.
Entries of 0.0 indicate a value $ < 0.05$. }
\label{tab:backgroundTableMuon_syst}
\centering
\small\addtolength{\tabcolsep}{+2pt}
\resizebox{\columnwidth}{!}{
\begin{tabular}{c r@{ $\pm$ }r r@{ $\pm$ }r r@{ $\pm$ }r r@{ $\pm$ }r r@{ $\pm$ }r r@{ $\pm$ }r}
\hline
\hline
\mmumu [GeV]& \multicolumn{2}{c}{\phantom{00}70-110}& \multicolumn{2}{c}{\phantom{00}110-130}& \multicolumn{2}{c}{130-150}& \multicolumn{2}{c}{150-170}& \multicolumn{2}{c}{170-200}\\
\hline
\zgstar   &7546.7 & 7.1  & 98.4 & 3.1  & 33.4 & 1.1  & 17.2 & 0.6  & 12.8 & 0.5  \\
\ttbar    &   6.0 & 0.6  &  2.4 & 0.3  &  1.7 & 0.2  &  1.2 & 0.1  &  1.2 & 0.1  \\
Diboson   &  10.0 & 0.5  &  0.8 & 0.1  &  0.6 & 0.0  &  0.5 & 0.0  &  0.4 & 0.0  \\
\wpjet    &   0.3 & 0.2  &  0.0 & 0.0  &  0.0 & 0.0  &  0.0 & 0.0  &  0.0 & 0.0  \\
QCD       &   0.1 & 0.0  &  0.0 & 0.0  &  0.0 & 0.0  &  0.0 & 0.0  &  0.0 & 0.0  \\
\hline
Total     &7563.0 & 7.2  &101.6 & 3.1  & 35.7 & 1.2  & 18.9 & 0.7  & 14.4 & 0.5  \\
\hline
Data  &  \multicolumn{2}{c}{7563}  &  \multicolumn{2}{c}{101}  &  \multicolumn{2}{c}{41}  &  \multicolumn{2}{c}{11}  &  \multicolumn{2}{c}{11}  \\
\hline
\hline
\mmumu [GeV] & \multicolumn{2}{c}{\phantom{000}200-240}& \multicolumn{2}{c}{\phantom{00}240-300}& \multicolumn{2}{c}{300-400}& \multicolumn{2}{c}{400-800}& \multicolumn{2}{c}{\phantom{00}800-2000 } \\
\hline
\zgstar &  7.8 & 0.3  &  5.1 & 0.2  &  2.5 & 0.1  &  1.3 & 0.1  &  0.1 & 0.0  \\
\ttbar  &  1.0 & 0.1  &  0.7 & 0.1  &  0.4 & 0.0  &  0.1 & 0.0  &  0.0 & 0.0  \\
Diboson &  0.3 & 0.0  &  0.2 & 0.0  &  0.2 & 0.0  &  0.1 & 0.0  &  0.0 & 0.0  \\
\wpjet  &  0.0 & 0.0  &  0.0 & 0.0  &  0.0 & 0.0  &  0.0 & 0.0  &  0.0 & 0.0  \\
QCD     &  0.0 & 0.0  &  0.0 & 0.0  &  0.0 & 0.0  &  0.0 & 0.0  &  0.0 & 0.0  \\
\hline
Total   &  9.1 & 0.4  &  6.0 & 0.2  &  3.0 & 0.1  &  1.5 & 0.1  &  0.1 & 0.0  \\
\hline
Data  &  \multicolumn{2}{c}{7}  &  \multicolumn{2}{c}{6}  &  \multicolumn{2}{c}{2}  &  \multicolumn{2}{c}{1}  &  \multicolumn{2}{c}{0}   \\
\hline
\hline
\end{tabular}
}
\end{table}

 Given the absence of a signal, an upper limit on the number of \zp\ events is determined at the 95\% confidence level (C.L.)
using a Bayesian approach~\cite{bayesianMethod}.
The invariant mass distribution of the data is compared to templates of the expected backgrounds and varying amounts of
signal at varying pole masses in the 0.13-2.0~TeV range, a technique used in Ref.~\cite{CDF:Zpmumu2fb}. 
A likelihood function is defined as the product of the Poisson
 probabilities over all mass bins in the search region, where the Poisson probability in each bin is evaluated for the observed
 number of data events given the expectation from the template. The total acceptance for signal as a function of mass is propagated into the expectation. 
For each \zp\ pole mass, a uniform prior in the \zp\ cross section is used.


The normalization procedure described above makes this analysis insensitive to the uncertainty on the
  integrated luminosity as well as other mass-independent systematic uncertainties. 
Mass-depen\-dent systematic uncertainties are incorporated as nuisance parameters whose variation is integrated over in the computation of the
 likelihood function~\cite{bayesianMethod}. 
The relevant systematic uncertainties are reconstruction efficiency, QCD and weak K-factors, PDF and resolution uncertainties. 
These uncertainties are correlated across all bins in the search region, 
and they are correlated between signal and background except for the weak K-factor which is only applied to the Drell-Yan background.
In addition, there is an uncertainty on the QCD component of the background for the electron channel.  

 The uncertainties on the mass-dependent nuisance parameters are as follows: since the total background is normalized to the data in the region of the 
 $Z\to\ll$ mass peak, the residual systematic uncertainties are small at low mass and grow at high mass. 
 The dominant uncertainties are of a theoretical nature. 
 The uncertainty on the cross sections due to PDF variation is 6\% (8\%) at 1~TeV for \zp\ (\zstar) production, for both channels.  
   The uncertainties on the QCD and weak K-factors are 3\% and 4.5\% respectively for both channels.
The uncertainty in the weak K-factor includes the
effects of neglecting real boson emission, the difference in the
electroweak scheme definition between \pythia\ and \horace, and higher
order electroweak and $\mathcal{O}(\alpha\alpha_s)$ corrections.
 Finally, an uncertainty of 5\%, due to the uncertainty
 on the $Z/\gamma^*$ cross section in the normalization region, as well as a 1\% statistical error on the data in the normalization region, are applied.  

 On the experimental side, the systematic effects are as follows. 
In the electron channel, the calorimeter resolution is dominated at large transverse energy by a constant term which is 1.1\% in the barrel and 1.8\% in the endcaps
with a small uncertainty.
The simulation was adjusted to reproduce this resolution at high energy and the uncertainty on it has a negligible effect.
The calorimeter energy calibration uncertainty is between 0.5\% and 1.5\% depending on transverse momentum and pseudorapidity. 
The non-linearity of the calorimeter response is negligible according to test beam data and Monte Carlo studies~\cite{Aharrouche:2006nf}.  
The uncertainty on the energy calibration has minimal impact on the sensitivity of the search, 
since its main effect is a shift of a potential peak in dilepton mass without change of the line-shape. 
 No source  of efficiency variation for electron reconstruction and identification at high \pt\ has been found. 
 
\begin{table}[!t]
\caption{Summary of systematic uncertainties on the expected numbers of events at $\mll=1$~TeV. 
NA indicates that the uncertainty is not applicable, and ``-'' denotes a negligible entry.}
\label{tab:systematicSummary}
\centering
\addtolength{\tabcolsep}{+1pt}
\resizebox{\columnwidth}{!}{
\begin{tabular}{lcccc}
\hline
\hline
Source        & \multicolumn{2}{c}{dielectrons} & \multicolumn{2}{c}{dimuons} \\
              & \zp\ signal & background        & \zp\ signal & background \\
\hline
Normalization & 5\%   & 5\%                     & 5\%   & 5\% \\
PDFs          & 6\%   & 6\%                     & 6\%   & 6\% \\
QCD K-factor  & 3\%   & 3\%                     & 3\%   & 3\% \\
Weak K-factor & NA    & 4.5\%                   & NA    & 4.5\% \\
Efficiency    & -     & -                       & 3\%   & 3\% \\
Resolution    & -     & -                       & 3\%   & 3\% \\
\hline
 Total        & 9.4\% & 9.5\%                   & 9.4\% & 10.4\%\\
\hline
\hline
\end{tabular}
}
\end{table}

 For the muon channel, 
 the combined uncertainty on the trigger and reconstruction efficiency is estimated to be 3\% at 1~TeV. 
 This uncertainty is dominated by the rate of muon bremsstrahlung in the calorimeter which may interfere with reconstruction in the muon spectrometer. 
 The uncertainty on the resolution due to residual misalignments in the muon spectrometer propagates to a change in the observed
  width of \zp/\zstar\ line-shape, and affects the sensitivity by 3\%. 
  The muon momentum scale is calibrated with a statistical precision of 0.2\% using the  $Z\to\ll$ mass peak. 
As with the electron channel, the momentum calibration uncertainty has negligible
 impact in the muon channel search. The systematic uncertainties are summarized in Table~\ref{tab:systematicSummary}. 


The limit on the number of \zp\ events produced is converted into a limit on cross section times branching ratio $\xbr (\zpll)$ 
by scaling with the observed number of \z\ boson events and the known value of $\xbr (Z \to \ll)$.
The expected exclusion limits are determined using simulated pseudo-experiments containing only Standard Model processes
by evaluating the 95\% C.L. upper limits for each pseudo-experiment for each fixed value of $M_{\zp}$.
The median of the distribution of limits is chosen to represent the expected limit. The ensemble of limits
 is also used to find the  68\% and 95\% envelope of the expected limits as a function of $M_{\zp}$.

\begin{figure}[!t]
  \centering
  \includegraphics[width=\columnwidth]{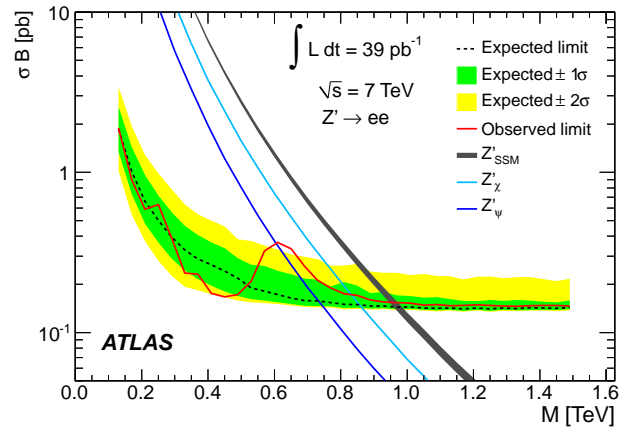}
  \caption{Expected and observed 95\% C.L. limits on \xbr\
           and expected \xbr\ for \zpssm\ production and the two \esix-motivated \zp\ models with lowest and highest \xbr\ 
	   for the dielectron channel. 
           The thickness of the SSM theory curve represents the theoretical uncertainty and holds for the other theory curves.
}
  \label{fig:eelimit_res}
\end{figure}

\begin{figure}[!htb]
  \centering
  \includegraphics[width=\columnwidth]{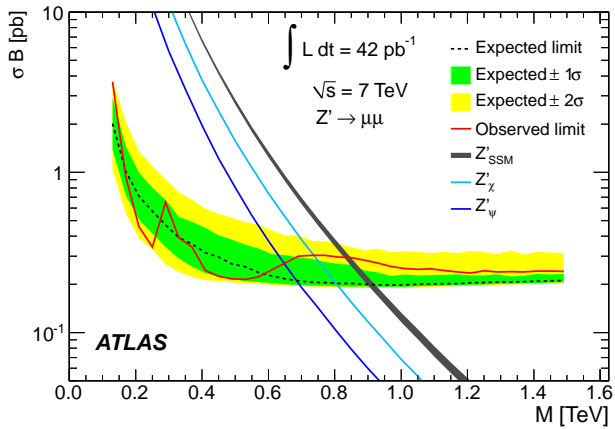}
  \caption{Expected and observed 95\% C.L. limits on \xbr\
           and expected \xbr\ for \zpssm\ production and the two \esix-motivated \zp\ models with lowest and highest \xbr\ 
	   for the dimuon channel. 
           The thickness of the SSM theory curve represents the theoretical uncertainty and holds for the other theory curves.
}
  \label{fig:mumulimit_res}
\end{figure}

\begin{figure}[!htb]
  \centering
  \includegraphics[width=\columnwidth]{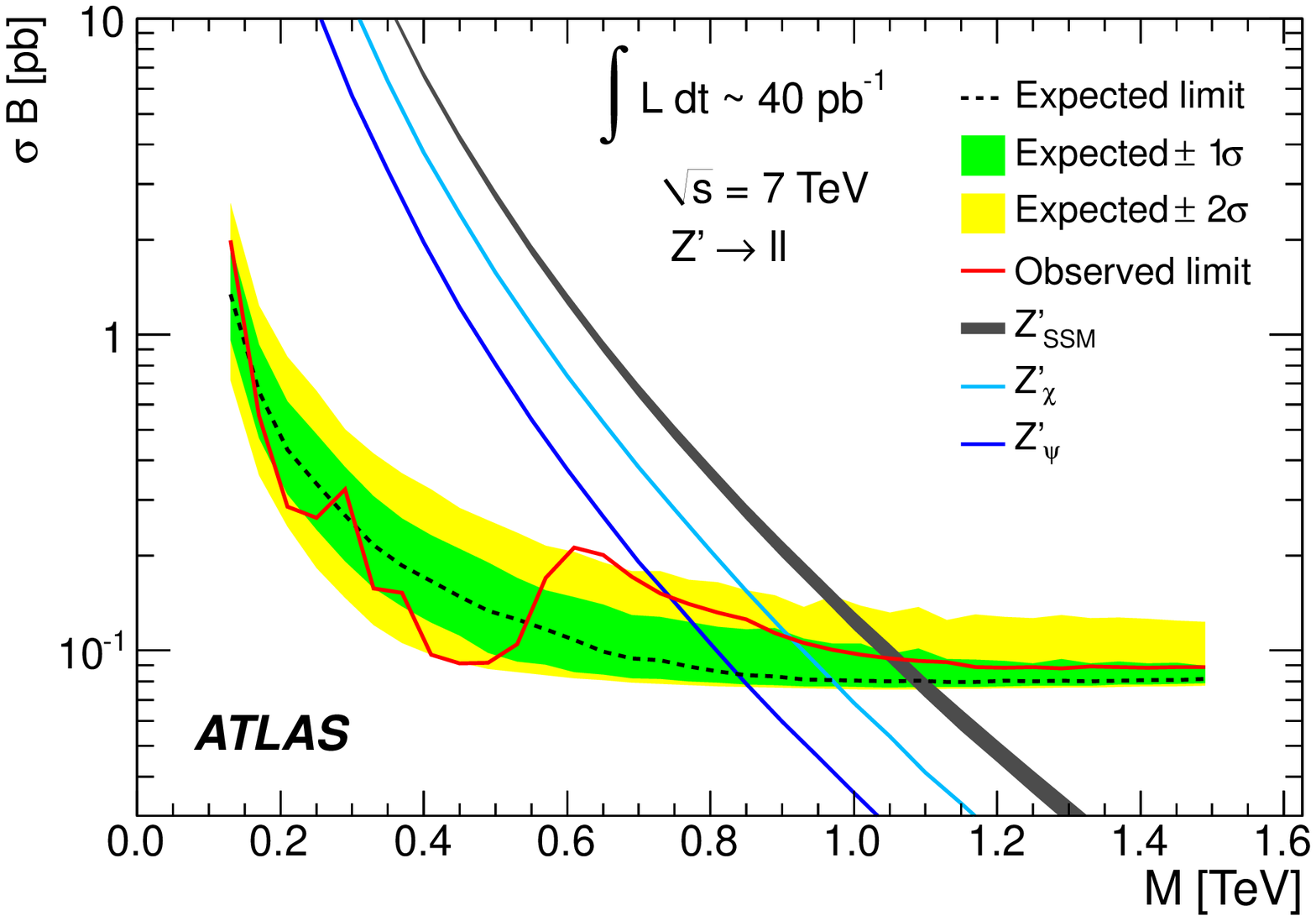}
  \caption{Expected and observed 95\% C.L. limits on \xbr\
           and expected \xbr\ for \zpssm\ production and the two \esix-motivated \zp\ models with lowest and highest \xbr\ 
	   for the combination of the electron and muon channels.
           The thickness of the \zpssm\ theory curve represents the theoretical uncertainty and holds for the other theory curves.
}
  \label{fig:combinedlimit_res}
\end{figure}

Figure~\ref{fig:eelimit_res} shows for the dielectron channel the 95\% C.L. observed and expected exclusion limits on \xbr.
It also shows the theoretical cross section times branching ratio for the \zpssm\ and for the lowest and highest \xbr\ of \esix-motivated \zp\ models.
Similarly, Figure~\ref{fig:mumulimit_res} shows the same results in the case of the dimuon channel.
 Figure~\ref{fig:combinedlimit_res} shows the 95\% C.L. exclusion limit on \xbr\ for the combination of the electron and muon channels. 
 The combination is performed by defining the likelihood function in terms of the total number of \zp\ events produced in both channels. 

In the three cases (dielectron, dimuon and combined channels), the 95\% C.L. \xbr\ limit is used to set mass limits for each of the considered models.
Mass limits obtained for the \zpssm\ are displayed in Table~\ref{tab:limits} together with the corresponding \xbr\ limit.
The combined mass limit for the \zpssm\ is 1.048~TeV (observed) and 1.088~TeV (expected). 
The combined mass limits on the \esix-motivated models are given in Table~\ref{e6massLimits}.
The limits on the \esix-motivated \zpI\ and \zpsq\ are 0.842~TeV and 0.871~TeV, more stringent than the previous highest limits~\cite{CDF:Zpmumu}.
\begin{table}[!hbt]
\caption{\ee, \mumu\ and combined 95\% C.L. mass and \xbr\ limits on \zpssm.
}
\label{tab:limits}
\begin{center}
\resizebox{\columnwidth}{!}{
\begin{tabular}{l|cc|cc}
\hline
\hline
             &  \multicolumn{2}{c|}{Observed limit} & \multicolumn{2}{c}{Expected limit} \\
                   & mass [TeV] & \xbr\ [pb]        & mass [TeV] & \xbr\ [pb]       \\
\hline			 	 	 	 	  	 
$\zpssm \to \ee$   & 0.957      & 0.155             & 0.967      & 0.145            \\
$\zpssm \to \mumu$ & 0.834      & 0.297             & 0.900      & 0.201            \\
$\zpssm \to \ll$   & 1.048      & 0.094             & 1.088      & 0.081            \\
\hline
\hline
\end{tabular}
}
\end{center}
\end{table}
\begin{table}[!hbt]
\caption{Combined mass limits at 95\% C.L. on the \esix-motivated \zp\ models.
}
\label{e6massLimits}
\begin{center}
\resizebox{\columnwidth}{!}{
\begin{tabular}{l|cccccc}
\hline
\hline
Model            & \zppsi & \zpN  & \zpeta & \zpI  & \zpsq & \zpchi  \\
\hline			 	 	 	 	  	 
Mass limit [TeV] & 0.738  & 0.763 & 0.771  & 0.842 & 0.871 & 0.900   \\
\hline
\hline
\end{tabular}
}
\end{center}
\end{table}

Although the lepton decay angular distributions are not the same for \zp\ and \zstar\ bosons, we found the difference in  geometrical acceptance to be negligible 
for boson pole masses above 750~GeV. 
The same procedure as for the \zp\ is used to calculate a limit on $\xbr (\zsll )$ and on the \zstar\ mass in each channel and for their combination.
The results are displayed in Table~\ref{tab:limits_zstar}.
The combined mass limit for the \zstar\ boson is 1.152~TeV (observed) and 1.185~TeV (expected). 
This is the first direct limit on this particle.
\begin{table}[!t]
\caption{\ee, \mumu\ and combined 95\% C.L. mass and \xbr\ limits on \zstar\ production.
}
\label{tab:limits_zstar}
\begin{center}
\resizebox{\columnwidth}{!}{
\begin{tabular}{l|cc|cc}
\hline
\hline
             &  \multicolumn{2}{c|}{Observed limit} & \multicolumn{2}{c}{Expected limit} \\
                   & mass [TeV] & \xbr\ [pb]        & mass [TeV] & \xbr\ [pb]       \\
\hline			 	 	 	 	  	 
\zsee              & 1.058      & 0.149             & 1.062      & 0.143            \\
\zsmumu            & 0.946      & 0.265             & 0.995      & 0.199            \\
\zsll              & 1.152      & 0.089             & 1.185      & 0.080            \\
\hline
\hline
\end{tabular}
}
\end{center}
\end{table}


In conclusion, the ATLAS detector has been used to search for narrow resonances in the invariant mass spectrum above 110~GeV of \ee\ and \mumu\
final states with $\sim 40$~\ipb\ of proton-proton data. No evidence for such a resonance is found.
Limits on the cross section times branching ratio $\xbr (\zpll )$ are calculated,
as well as mass limits on the \zpssm\  (1.048~TeV), the \zstar\ (1.152~TeV) and \esix-motivated \zp\ bosons (in the range 0.738$-$0.900~TeV).
For certain \esix-motivated models, these limits are more stringent than the corresponding limits from the Tevatron.\\



We wish to thank CERN for the efficient commissioning and operation of the LHC during this initial high-energy data-taking period as well as the support staff from our institutions without whom ATLAS could not be operated efficiently.

We acknowledge the support of ANPCyT, Argentina; YerPhI, Armenia; ARC, Australia; BMWF, Austria; ANAS, Azerbaijan; SSTC, Belarus; CNPq and FAPESP, Brazil; NSERC, NRC and CFI, Canada; CERN; CONICYT, Chile; CAS, MOST and NSFC, China; COLCIENCIAS, Colombia; MSMT CR, MPO CR and VSC CR, Czech Republic; DNRF, DNSRC and Lundbeck Foundation, Denmark; ARTEMIS, European Union; IN2P3-CNRS, CEA-DSM/IRFU, France; GNAS, Georgia; BMBF, DFG, HGF, MPG and AvH Foundation, Germany; GSRT, Greece; ISF, MINERVA, GIF, DIP and Benoziyo Center, Israel; INFN, Italy; MEXT and JSPS, Japan; CNRST, Morocco; FOM and NWO, Netherlands; RCN, Norway;  MNiSW, Poland; GRICES and FCT, Portugal;  MERYS (MECTS), Romania;  MES of Russia and ROSATOM, Russian Federation; JINR; MSTD, Serbia; MSSR, Slovakia; ARRS and MVZT, Slovenia; DST/NRF, South Africa; MICINN, Spain; SRC and Wallenberg Foundation, Sweden; SER,  SNSF and Cantons of Bern and Geneva, Switzerland;  NSC, Taiwan; TAEK, Turkey; STFC, the Royal Society and Leverhulme Trust, United Kingdom; DOE and NSF, United States of America.  

The crucial computing support from all WLCG partners is acknowledged gratefully, in particular from CERN and the ATLAS Tier-1 facilities at TRIUMF (Canada), NDGF (Denmark, Norway, Sweden), CC-IN2P3 (France), KIT/GridKA (Germany), INFN-CNAF (Italy), NL-T1 (Netherlands), PIC (Spain), ASGC (Taiwan), RAL (UK) and BNL (USA) and in the Tier-2 facilities worldwide.


\bibliography{Zp}{}

\begin{widetext}
\clearpage

\begin{flushleft}
{\Large The ATLAS Collaboration}

\bigskip

G.~Aad$^{\rm 48}$,
B.~Abbott$^{\rm 111}$,
J.~Abdallah$^{\rm 11}$,
A.A.~Abdelalim$^{\rm 49}$,
A.~Abdesselam$^{\rm 118}$,
O.~Abdinov$^{\rm 10}$,
B.~Abi$^{\rm 112}$,
M.~Abolins$^{\rm 88}$,
H.~Abramowicz$^{\rm 153}$,
H.~Abreu$^{\rm 115}$
% ...
\bigskip

% ...
$^{10}$ Institute of Physics, Azerbaijan Academy of Sciences, Baku, Azerbaijan\\
$^{11}$ Institut de F\'isica d'Altes Energies and Universitat Aut\`onoma  de Barcelona and ICREA, Barcelona, Spain\\
% ...
$^{48}$ Fakult\"{a}t f\"{u}r Mathematik und Physik, Albert-Ludwigs-Universit\"{a}t, Freiburg i.Br., Germany\\
$^{49}$ Section de Physique, Universit\'e de Gen\`eve, Geneva, Switzerland\\
% ...
$^{88}$ Department of Physics and Astronomy, Michigan State University, East Lansing MI, United States of America\\
% ...
$^{111}$ Homer L. Dodge Department of Physics and Astronomy, University of Oklahoma, Norman OK, United States of America\\
$^{112}$ Department of Physics, Oklahoma State University, Stillwater OK, United States of America\\
% ...
$^{115}$ LAL, Univ. Paris-Sud and CNRS/IN2P3, Orsay, France\\
% ...
$^{118}$ Department of Physics, Oxford University, Oxford, United Kingdom\\
% ...
$^{153}$ Raymond and Beverly Sackler School of Physics and Astronomy, Tel Aviv University, Tel Aviv, Israel\\
% ...
\end{flushleft}

\end{widetext}

\end{document}